\documentclass[conference]{IEEEtran}
\IEEEoverridecommandlockouts
\usepackage{cite}
\usepackage{amsmath,amssymb,amsfonts}
\usepackage{algorithmic}
\usepackage{graphicx}
\usepackage{textcomp}
\usepackage{xcolor}
\usepackage{multirow}
\usepackage{booktabs}
\usepackage{float}
\usepackage{graphicx}
\usepackage[super]{nth}

\def\BibTeX{{\rm B\kern-.05em{\sc i\kern-.025em b}\kern-.08em
    T\kern-.1667em\lower.7ex\hbox{E}\kern-.125emX}}
\begin{document}

\title{Subject-Independent Classification of Brain Signals using Skip Connections
\footnote{{\thanks{This work was partly supported by Institute for Information \& Communications Technology Planning \& Evaluation (IITP) grant funded by the Korea government (MSIT) (No. 2017-0-00451, Development of BCI based Brain and Cognitive Computing Technology for Recognizing User’s Intentions using Deep Learning; No.2021-0-02068, Artificial Intelligence Innovation Hub).}
}}
}

\author{\IEEEauthorblockN{Soowon Kim}
\IEEEauthorblockA{\textit{Dept. Artificial Intelligence} \\
\textit{Korea University} \\
Seoul, Republic of Korea \\
soowon\_kim@korea.ac.kr} \\

\IEEEauthorblockN{Young-Eun Lee}
\IEEEauthorblockA{\textit{Dept. Brain and Cognitive Engineering} \\
\textit{Korea University} \\
Seoul, Republic of Korea \\
ye\_lee@korea.ac.kr} \\

\and
\IEEEauthorblockN{Ji-Won Lee}
\IEEEauthorblockA{\textit{Dept. Artificial Intelligence} \\
\textit{Korea University} \\
Seoul, Republic of Korea \\
jiwon\_lee@korea.ac.kr} \\

\IEEEauthorblockN{Seo-Hyun Lee}
\IEEEauthorblockA{\textit{Dept. Brain and Cognitive Engineering} \\
\textit{Korea University} \\
Seoul, Republic of Korea \\
seohyunlee@korea.ac.kr} \\


}

\IEEEoverridecommandlockouts
\IEEEpubid{\makebox[\columnwidth]{978-1-6654-6444-4/23/\$31.00~\copyright2023 IEEE \hfill}
           \hspace{\columnsep}\makebox[\columnwidth]{ }}

\maketitle

\begin{abstract}

Untapped potential for new forms of human-to-human communication can be found in the active research field of studies on the decoding of brain signals of human speech. A brain-computer interface system can be implemented using electroencephalogram signals because it poses more less clinical risk and can be acquired using portable instruments. One of the most interesting tasks for the brain-computer interface system is decoding words from the raw electroencephalogram signals. Before a brain-computer interface may be used by a new user, current electroencephalogram-based brain-computer interface research typically necessitates a subject-specific adaption stage. In contrast, the subject-independent situation is one that is highly desired since it allows a well-trained model to be applied to new users with little or no precalibration. The emphasis is on creating an efficient decoder that may be employed adaptively in subject-independent circumstances in light of this crucial characteristic. 
Our proposal is to explicitly apply skip connections between convolutional layers to enable the flow of mutual information between layers. To do this, we add skip connections between layers, allowing the mutual information to flow throughout the layers. The output of the encoder is then passed through the fully-connected layer to finally represent the probabilities of the 13 classes. In this study, overt speech was used to record the electroencephalogram data of 16 participants. The results show that when the skip connection is present, the classification performance improves notably.
\end{abstract}

\begin{small}
\end{small}

\section{INTRODUCTION}
Brain signals carry information about human behavior, imagery or mental and physical states \cite{zhang2017hybrid}, which is useful for deciphering intents. By analyzing a user's brain activity, brain-computer interface (BCI) technology can generate external commands that can be utilized to control the surroundings\cite{castermans2011optimizing, thung2018conversion, jeong2020decoding}. Through the use of their intents, which are translated from brain signals, users of BCI can control prostheses, offering an additional means of communication between people and external machines\cite{won2017motion}. A recent area of BCI research, brain-to-speech (BTS), aims to produce spoken words from brain signals. The goal of BTS systems is to decipher the speech-related intents from brain activity and then provide orders for communication, in contrast to other communication techniques like event-related potential spellers. Brain-to-brain systems may be a means of intuitive communication as speech is the most common mode of communication\cite{huth2016natural}. We hypothesize that there must be a meaningful brain activation that might encode a substantial aspect of the speech because it has been demonstrated that it is possible to reconstruct speech from brain signals of spoken speech\cite{jeong2020brain}. EEG signal analysis methods have seen a great deal of development, with promising results \cite{yadava2017analysis, roy2019deep, ewen2019conceptual, lee2019connectivity, suk2014predicting, lee2018high}. A brief calibration session is required before a BCI system may be utilized by a new user because the majority of recent studies concentrate on the subject-dependent scenario in which training and test data are from the same subject \cite{liu2014real, sadiq2020identification, jeong2020brain}. This labor- and time-intensive calibration procedure must be carried out for every new subject and usage. The subject-independent scenario, in contrast, is little studied but is greatly sought to enhance the user-experience. In this scenario, a BCI system is trained using data from seen individuals while being applied straight to new users without pre-adaptation. Exploring the subject-independent scenario, however, is quite challenging. EEG signals change significantly even between recording sessions of the same user under the same experimental paradigm and show strong subject-to-subject variability \cite{kwak2015lower}. The majority of techniques rely on conventional machine learning techniques and manually created features, making them one of the few subject-independent investigations. \cite{khalid2016epileptic} extracts and categorizes information from each subject's EEG data using a pair of Linear Discriminant Analysis (LDA) and Common Spatial Patterns (CSP) algorithms. An ensemble classifier is created by combining many classifiers using l1 regularized regression. Due to the handcrafted features' and conventional learning algorithms' limited capabilities, these methods are unable to deliver performance that is sufficient. Deep learning techniques have recently made impressive strides, showing promise for addressing complex cross-subject scenarios \cite{lee2017network, lee2020decoding}. However, only a small number of successful deep learning studies have shown strong generalization capabilities from known subjects to new ones \cite{waytowich2018compact, bhatti2019soft, lee2020neural, kim2019subject, kwon2019subject}. Therefore, deep learning based models for EEG signal decoding still remains challenging tasks \cite{lee2019eeg}. 

Skip connection is a common method for enhancing the performance and convergence of deep neural networks. It works by propagating a linear component through the layers of the neural network, which is thought to ease the difficulties in optimization caused by non-linearity \cite{liu2020rethinking, he2016deep}. As the name implies, Skip Connections (or Shortcut Connections) omit some neural network layers and provide the result of one layer as the input to subsequent levels. Skip Connections were developed to address various issues in various architectures. Skip connections allowed us to address the degradation issue in the case of ResNets, while it ensured feature reusability in the case of DenseNets \cite{huang2017densely}. In the sections below, we'll go into further detail on each one.

In this article, we introduce an unique subject-independent EEG data analysis method based on convolutional neural network that is taking advantage of skip connection. Through multiple skip connections the flow of mutual information is preserved. The encoding of EEG signals is accomplished using convolutional neural networks.
\begin{figure}[t]
\centering
    \includegraphics[width=\linewidth]{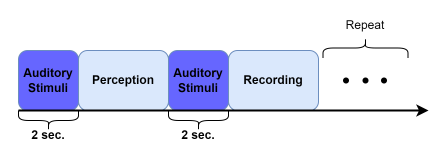}
    \caption{Experimental paradigm for recording EEG signals from overt speech}
    \label{fig_para}
\end{figure}

\begin{figure}[t]
\centering
    \includegraphics[width=7cm]{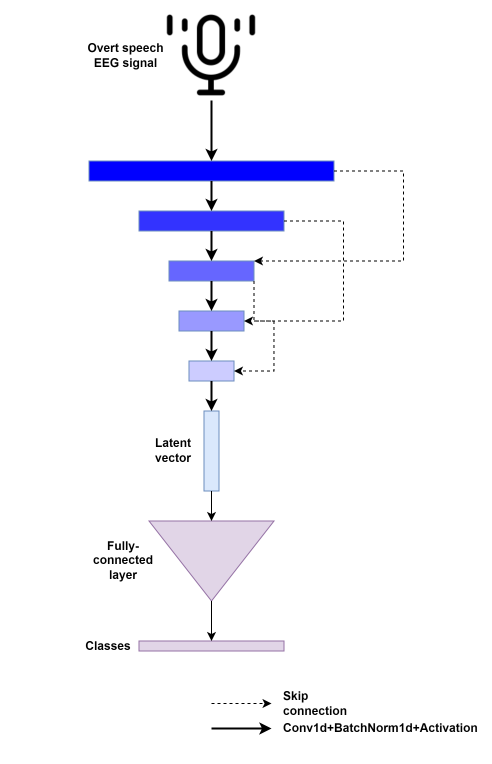}
    \caption{Overall architecture of the model used in this work. Waveform of the overt speech EEG signals is used as an input to the model with 64 channels. Each block performas 1-dimensional convolution, 1-dimensional batch normalization, and activation (ELU). To enable skip connection throughout the intermediate layers, maxpooling is applied to match the dimensions. After extracting the latent vector from the CNN, fully-connected layer is applied to output the final classification probabilities.}
    \label{fig_arch}
\end{figure}
\section{Materials and methods}

\subsection{Subjects}
The experimental protocol was designed to record EEG signals while performing overt speech following the perception stage with auditory stimuli in between. Sixteen subjects participated in the study. The study was approved by the Korea University Institutional Review Board [KUIRB-2019-0143-01] and was conducted in accordance with the Declaration of Helsinki. Informed consent was obtained from all subjects. All subjects are asked to fill out the questionnaire before and after the experiment to check their physiological and mental conditions and evaluate the experimental paradigms.

\subsection{Experimental Protocol and Paradigm}
We recorded EEG signals from scalp during overt speech through 64 channels. First, an auditory stimuli is given to the subject which is a beeping sound. Then, the perception is performed which is a hearing of the given word with visual perception as well through the monitor. With another auditory stimuli, the corresponding overt speech is performed. EEG signals from both stages are collected. In total, more than 1,400 trials are performed in each subject The experimental paradigm is described in Fig. \ref{fig_para}. Thirteen classes are introduced, with labels ranging from 0 to 12. These classes include ``ambulance," ``clock," ``hello," ``help me," ``light," ``pain," ``stop," ``thank you," ``toilet," ``TV," ``water," and ``yes," and a silent phase.

\begin{table}[]
\centering
\caption{Classification performance of deep convolutional neural networks with and without skip connections}
\label{tab:my-table}
\resizebox{\columnwidth}{!}{%
\begin{tabular}{lllll}
\hline
\textbf{Subject}                 & \textbf{Accuracy (\%)} & \textbf{F1-score (\%)} & \textbf{Precision (\%)} & \textbf{Recall (\%)} \\ \hline
\textbf{Proposed method}    & 98.69           & 98.70           & 98.74            & 98.69         \\
\textbf{Without skip connection} & 80.24           & 80.64           & 83.38              & 80.25               \\ \hline
\end{tabular}%
}
\end{table}
\subsection{Preprocessing}
From the beginning of each trial, the EEG signal was segmented into 1.5 seconds and sampled at 1,000 Hz. A fifth Butterworth filter was used to preprocess the EEG data in the high-gamma region of 0.5-120 Hz, and baseline was corrected by deducting the average of 500 ms before to the start of each trial. We chose the channels (AF3, F3, F5, FC3, FC5, T7, C5, TP7, CP5, and P5) that are in the Broca and Wernicke's areas. Using independent component analysis and references from EOG and EMG, we performed artifact reduction techniques for the muscle activity around the mouth. Using OpenBMI Toolbox \cite{lee2019eeg}, BBCI Toolbox \cite{krepki2007berlin}, and EEGLAB \cite{delorme2004eeglab}, all data processing operations were carried out in Python and Matlab.

\subsection{Architecture}
The proposed classification framework consists of convolution layers and skip connection to extract time-spectral-spatial information, as shown in Fig. \ref{fig_arch}. The raw EEG signals as a waveform were used as an input to the model. To capture various EEG features such as spectral, spatial, and temporal information, we used deep convolutional network. In each encoding block, 1-dimensional convolutional layer is followed by 1-dimensional batch normalization, activation and dropout. The probability of the dropout was 0.5. In total, 5 encoding blocks are used to produce the final latent representation of the raw EEG signals. We used the exponential linear unit (ELU) as the activation function. 64 channels are maintained throughout each encoding block.

The classifier (Fully-connected layer) is then applied with 3 linear, ELU activation and dropout combination. 128 hidden units are used. The output for classification is set to 13 classes with the input as raw signals (C × T). For training, we applied the mean squared error loss. For each condition, 2000 epoch training and 5-fold cross-validation were used to conduct the evaluation. The likelihood of chance level for this experiment was \(5\%\) due to the 13 classes present.

\begin{figure}
      \label{fig:fig1}
      \centering
      \includegraphics[width=.8\linewidth]{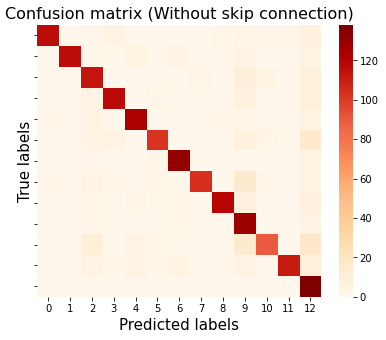}
      \includegraphics[width=.8\linewidth]{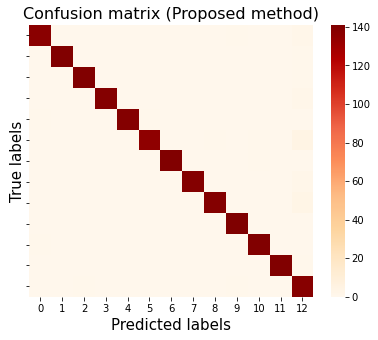}
      
\caption{The confusion matrix for the model comparing the proposed method with no skip connection. The true labels are on the vertical axis, while the predicted labels are on the horizontal axis. The upper matrix represent the results from the model without skip connection and the lower matrix represents the model with skip connection.}
\label{fig_cm_skip}
\end{figure}

\subsection{Model training}
The training session consists of more than 1,000 epochs. For every 10 epochs, we validated the model with a validation set, which is \(10\%\) randomly selected portion of the entire dataset. When the model's performance does not increase for more than 10 validation, we stop the training.

Adam was used for the training optimizer. The learning rate was \(0.0001\) which was fixed throughout the training without weight decay with betas \((b_1, b_2)\) with 0.9 and 0.999. A batch size of 128 was used for all experiments.

\section{Results and Discussion}
For 13 classes of overt speech, we created frameworks for deciphering speech-related EEG data. For overt speech, the performance of the model with and without skip connection was examined. As table \ref{tab:my-table} shown above, for 13 classes with skip connections, the average accuracy of overt speech was 98.69\%, compared to 80.24\% for the model without skip connections. F1-score, as well as precision and recall showed clear advantage, as shown in table \ref{tab:my-table}, indicating a better performance of the model using the skip connection compared to the model without skip connection. The confusion matrix for the model with skip connection is depicted in Fig. \ref{fig_cm_skip}.


\section{Conclusion}
In this study, we proposed a deep convolutional neural network for EEG decoding that applies skip connections in between convolutional layers to enable the flow of mutual information between layers. To do this, we add a skip connection between first, second and third convolutional layer, allowing the mutual information to flow from the input layer to the output layer. The output of the encoder is then passed through the fully-connected layer to finally represent the probabilities of the 13 classes. According to the findings, the performance was significantly enhanced when the subject-independent classification was carried out on seven subjects utilizing the skip connection. As a result, the method that was proposed for decoding brain activity with skip connections had the potential to be used with reliable BCI devices for any subject.

\bibliographystyle{IEEEtran}
\bibliography{REFERENCE}

\end{document}